\documentclass[aps,prb,twocolumn,superscriptaddress,preprintnumbers,amsmath,amssymb,floatfix,showpacs]{revtex4}

\usepackage{graphicx, epstopdf}
\usepackage{dcolumn}
\usepackage{bm}

\begin{document}

\title{Ultrafast Dynamic Metallization of Dielectric Nanofilms by Strong Single-Cycle Optical Fields}


\author{Maxim Durach}
\affiliation{
Department of Physics and Astronomy, Georgia State
University, Atlanta, Georgia 30303, USA}
\author{Anastasia Rusina}
\affiliation{
Department of Physics and Astronomy, Georgia State
University, Atlanta, Georgia 30303, USA}
\author{Matthias F. Kling}
\affiliation{Max-Planck-Institut f\"ur Quantenoptik,
Hans-Kopfermann-Stra{\ss}e 1, D-85748 Garching,
Germany}
\author{Mark I. Stockman}
\affiliation{
Department of Physics and Astronomy, Georgia State
University, Atlanta, Georgia 30303, USA
\\
E-mail: mstockman@gsu.edu; Homepage: http://www.phy-astr.gsu.edu/stockman
\email{mstockman@gsu.edu}
\homepage{http://www.phy-astr.gsu.edu/stockman}
}
\affiliation{Max-Planck-Institut f\"ur Quantenoptik,
Hans-Kopfermann-Stra{\ss}e 1, D-85748 Garching,
Germany}

\date{\today}

\begin{abstract}

We predict a dynamic metallization effect where an ultrafast (single-cycle) optical pulse with a $\lesssim 1$ V/\protect{\AA} field causes plasmonic metal-like behavior of a dielectric film with a few-nm thickness. This  manifests itself in plasmonic oscillations of polarization and a significant population of the conduction band evolving on a $\sim 1$ fs time scale. These phenomena are due a combination of both adiabatic (reversible) and diabatic (for practical purposes irreversible) pathways.

\end{abstract}


\maketitle

Latest advances in the ultrafast optics have recently attracted a great deal of attention. Ultrashort pulses have been successfully employed for monitoring and manipulation of electronic processes in atomic and molecular structures. \cite{Krausz_Ivanov_RevModPhys.81.163_2009_Attosecond_Review} Significant  efforts have been directed toward exploring the potential of ultrashort ($\sim 100$ as to $\sim 1-10$ fs in duration) pulses in application to condensed matter dynamics \cite{Murnane_et_al_PRL_97_113604_2006_Laser_Assisted_Photoelectric_Effect_from_Surfaces, Cavalieri_et_al_Nature_2007_Attosecond_Photoemission_from_Solids,   Murnane_Kapteyn_et_al_PRL_2009_Demagnetization_Dynamics, Murnane_PRA_2009_Dressed_Electronic_Processes_at_Surfaces, Murnane_et_al_Nature_Materials_2010_Ballistic_Thermal_Transport, Stockman_et_al_J_Phys_2009_PEEM, Reis_et_al_Nature_Phys_2011_HHG_from_ZnO_Crystal}, in particular, to plasmonic metal and dielectric nanostructures. \cite{Stockman_Hewageegana_APA_2007_CEP,  Stockman_Kling_Kleineberg_Krausz_Nature_Photonics_2007, Kling_et_al_Nature_Phot_2011_Dielectric_Sphere_Ultrafast_Photoemission}

We have recently predicted that dielectric nanofilms subjected to strong but sufficiently slow (adiabatic) electric fields undergo a reversible change resembling a quantum phase transition to a state that exhibits metallic optical properties. \cite{Stockman_et_al_PRL_2010_Metallization} We have called this phenomenon metallization. The minimum duration of such an adiabatic field depends exponentially on the thickness of the nanofilm and is in the range from $\gtrsim 10$ fs to $\sim 10$ ns for a film thickness from a few nm to $\sim 10$ nm.  \cite{Stockman_et_al_PRL_2010_Metallization}

Both from the fundamental point of view and for applications to ultrafast nanoelectronics, the metallization by much faster optical fields is of great interest. In this Letter, we theoretically predict a new effect that we call {\em dynamic metallization}, where a single-cycle optical pulse incident on a $\sim 2$ nm dielectric nanofilm with a normal polarization and a field $\lesssim 1$ V/\protect{\AA} causes a population of the conduction band and metal-like plasmonic polarization oscillations on the optical-period time scale. This effect is caused by both adiabatic (reversible) and diabatic (dissipative) excitation pathways involving band anticrossings and adiabatic evolution between them.

A comprehensive solution of the ultrafast electron dynamics in strong optical fields would require many-body quantum kinetics, rendering this problem extremely complicated. To simplify it, we rely on the fact that the characteristic inelastic electron-scattering time is on the order or greater than the surface plasmon decay time $\tau_n$, which is $\tau_n\gtrsim 10$ fs for metals -- see, e.g., Fig.\ 1 (a) in Ref.\ \onlinecite{Bergman_Stockman:2003_PRL_spaser}. Using ultrashort excitation pulses with duration $\tau\ll \tau_n$, we avoid  any significant effect of the electron inelastic scattering. This allows us to treat the electron dynamics as Hamiltonian. The evolution of the system in this case is convenient to describe by the density matrix 
\begin{equation}
\hat\rho(\mathbf r^\prime,\mathbf r; t)=\sum_{i\le i_F} \Psi_i(\mathbf r^\prime, t) \Psi_i^\ast(\mathbf r, t)~,
\label{rho}
\end{equation}
where $i_f$ denotes the Fermi-surface state, i.e., the highest occupied state for the zero-field Hamiltonian $\hat H_0$, and $\Psi_i(\mathbf r^\prime, t)$ are the one-electron wave functions. These satisfy the Schr\"odinger equation $i\hbar\dot \Psi_i=\hat H(t)\Psi_i$, where $\hat H(t)$ is the Hamiltonian depending on time $t$ due to the optical field, and the dot denotes the  derivative over $t$. 

Consider a thin nanofilm where the energy bands are split into subbands due to the quantum confinement in the direction normal to the film plane. We assume that, due to the material symmetry, the electron wave function can be factorized into normal and parallel to the film. Assuming a normal optical electric field  $\mathcal E=\mathcal E(t)$, the one-particle Hamiltonian of the transverse motion is $\hat H(\mathcal E) = \hat H_0 - \mathcal E \hat d$, where $\hat d$ is the dipole operator.

\begin{figure}
\includegraphics[width=.48\textwidth]{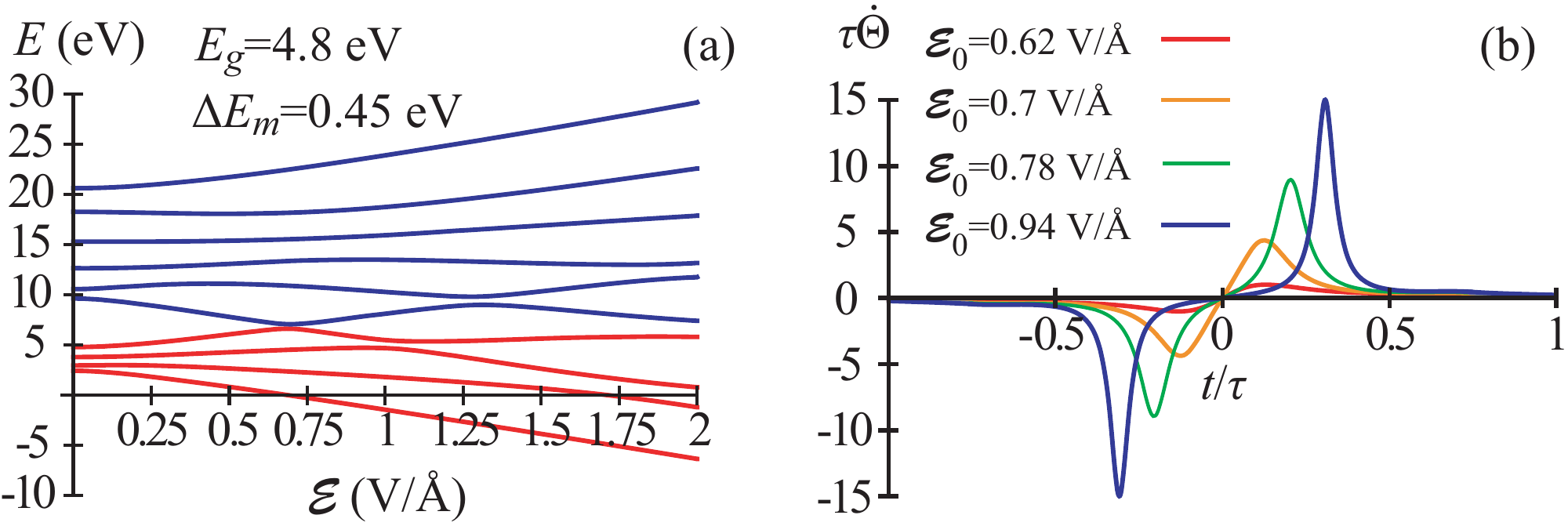}
\caption{
(a) Energy spectrum of the nanofilm as a function of the adiabatically applied electric field. The occupied valence subbands are shown in red, the empty conduction subbands are in blue. (b) The diabatic coupling matrix element $\dot \Theta_{ik}$ between band-edge subbands [see Eq.\ (\ref{nangle})] for different pulse amplitudes $\mathcal E_0$, as indicated on the panel. 
}
\label{adiabatic_levels.eps}
\end{figure}

Consider the adiabatic basis of states $\psi_i(\mathcal E)$ that diagonalize the instantaneous Hamiltonian, $\hat H(\mathcal E) \psi_i(\mathcal E) = E_i(\mathcal E)\psi_i(\mathcal E) $, where $E_i(\mathcal E)$ are the adiabatic energies. We employ the Kronig-Penney model for an insulator with $E_g=4.8$ eV band gap at the zero field (simulating diamond) whose adiabatic subband energies $E_i(\mathcal E)$ of the valence (red) and conduction (blue) bands are shown in Fig.\ \ref{adiabatic_levels.eps} (a) as functions of the applied field $\mathcal E$. All calculations are done for a $l=2$ nm thickness film. 

We expand wave functions $\Psi_i(t)$ in the adiabatic basis,  $\Psi_i(t)=\sum_j \mathrm{exp}[-i \varphi_j(t)] a_j^{(i)}(t)  \psi_j\left(\mathcal E(t)\right)$, where $a_j^{(i)}(t)$ are the expansion coefficients with the initial condition $a_j^{(i)}(0)=\delta_{ij}$, and  $\varphi_j(t)=-(1/\hbar)\int E_j \left(\mathcal E(t)\right) dt$ is the adiabatic phase. Then the Schr\"odinger equation becomes
\begin{eqnarray}
&&\dot a_j^{(i)}=-\sum_{k\ne j}\dot\Theta_{jk} \mathrm{exp}\left[-i\varphi_{jk}(t)\right] a_k^{(i)}~,~~~
\label{dota}
\\
&&\dot{\Theta}_{jk}\equiv -\dot{\mathcal E} d_{jk}\left(\mathcal E\right)/E_{jk}\left(\mathcal E\right)~,
\label{nangle}
\end{eqnarray}
where  the adiabatic dipole matrix elements, transition energies, and relative phases are $d_{jk}(\mathcal E)=\langle \psi_k(\mathcal E)|\hat d_0|\psi_j(\mathcal E)\rangle$ and $E_{jk}(\mathcal E)=E_j(\mathcal E)-E_k(\mathcal E)$, $\varphi_{jk}(t)=\varphi_{j}(t)-\varphi_{k}(t)$, correspondingly. 

Under the adiabatic conditions \cite{Stockman_et_al_PRL_2010_Metallization}, a strong electric field causes  the band gap $E_g$ to decrease. The valence and conduction bands experience  anticrossing at the metallization field  $\mathcal E_m=0.75$~V/\protect{\AA} with the anticrossing gap  $\Delta E_m=0.45$~eV -- see Fig.\  \ref{adiabatic_levels.eps} (a). If the field is increased adiabatically above $\mathcal E_m$, the electrons are adiabatically (reversibly) transferred to the conduction band (hole) states and in space across the film. This is the metallization transition where the optical properties of the nanofilm resemble those of a plasmonic metal. \cite{Stockman_et_al_PRL_2010_Metallization} If the field is adiabatically switched off, the system returns to its ground state. The condition of the adiabaticity is evident from Eqs.\ (\ref{dota})-(\ref{nangle}) and is $t_p\gg \hbar/\Delta E_m$, where $t_p$ is the time needed for the field $\mathcal E(t)$ to pass through the anticrossing,  in full agreement with Ref.\ \onlinecite{Stockman_et_al_PRL_2010_Metallization}.

In the opposite case of a fast diabatic passage of the anticrossing, the Schr\"odinger equation (\ref{dota}) for the valence and conduction band-edge subbands, $v$ and $c$, can be integrated yielding the population of the conduction band
\begin{equation}
n_c(t)=\mathrm{sin}^2 \Theta_{vc}\left[\mathcal E(t)\right]n/n_{sb}~,~~
\label{diabatic_excitation} 
\end{equation}
where $n$ is the electron density and $n_{sb}$ is the number of the occupied subbands, $n_{sb}=9$ in the present model. Such rapid fields, in contrast to the adiabatic case,  \cite{Stockman_et_al_PRL_2010_Metallization} do not induce the metal-like polarization: there is no spatial population transfer across the nanofilm.

\begin{figure}
\includegraphics[width=.48\textwidth]{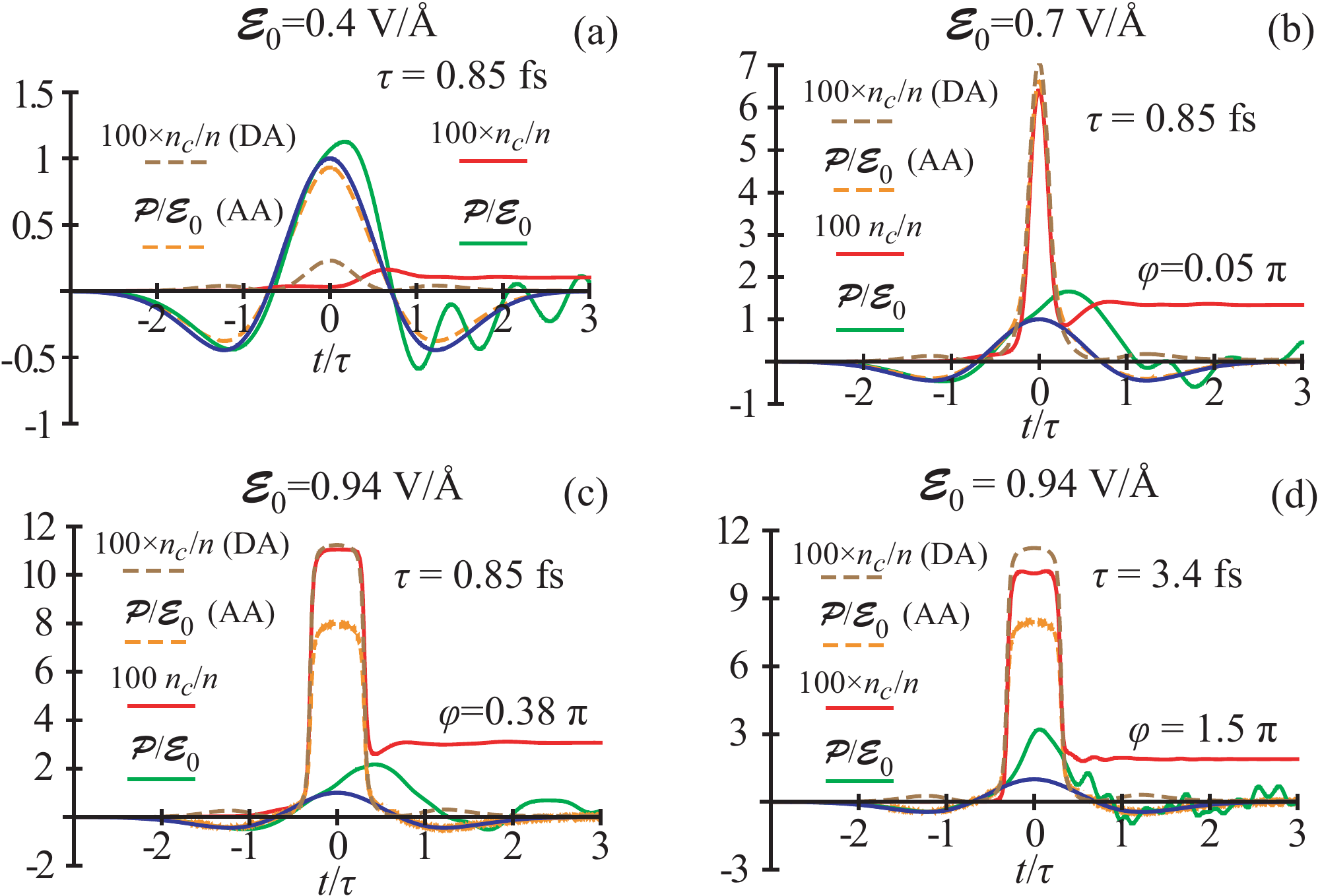}
\caption{
Polarization $\mathcal P$ and conduction-band population $n_c$ as functions of time  $t$ for various $\mathcal E_0$ and $\tau$. Normalized pulse field $\mathcal E(t)/\mathcal E_0$  is shown by blue line. Normalized population $n_c(t)/n$ (scaled $\times 100$)  is displayed by the red curve; the same in DA is shown by the dashed gray curve. Relative polarization $\mathcal P/\mathcal E_0$ is displayed by the green line; the same in AA is given by the dashed yellow line. The pulse length is $\tau=0.85$ fs for (a)-(c) and $\tau=3.4$ fs for (d).
}
\label{amplitude_dependence.eps}
\end{figure}

We consider the electron dynamics of a nanofilm subjected to an ultrafast field where both the adiabatic and diabatic processes contribute. The fastest dynamics is driven by single-cycle light pulses with duration $\tau\sim 1$~fs, which have recently been achieved. \cite{Goulielmakis_et_al_Single_Cycle_Nonlinear_Optics_Science_2008} Here, we model a single-cycle pulse $\mathcal E(t)$ by a waveform \cite{Rastunkov_Krainov_one_cycle_pulse_shape}
\begin{equation}
\mathcal E(t) = \mathcal E_0 e^{-u^2}\left(1-2u^2\right)~,~~~u\equiv t/\tau~,
\label{pulse_shape}
\end{equation}
where the amplitude is $\mathcal E_0$, and the pulse duration is $\tau$. The pulse integral is zero,  $\int_{-\infty}^{\infty}\mathcal E(t) dt=0$, as should be.

For such a pulse, in Fig.\ \ref{adiabatic_levels.eps} (b) we display the diabatic coupling matrix element $\dot\Theta_{ik}$ between  the valence and conduction band-edge subbands. Note that $\dot\Theta_{ik}\propto 1/\tau$. The peaks of the diabatic coupling element $\dot\Theta_{ik}$ are at the adiabatic metallization points (band-edge anticrossings), and they grow with the excitation-pulse amplitude.

Now let us turn to the dynamics of the system excited by an ultrashort pulse with $\tau=0.85$ fs (the mean frequency $\hbar\omega_0=2\hbar/\tau=1.55$ eV). This is illustrated in Figs.\ \ref{amplitude_dependence.eps} (a)-(c) where we show conduction band population $n_c$ and polarization $\mathcal P=\mathrm{Tr}\{\hat d\rho\}/V$, where $V$ is the nanofilm's volume, as functions of time $t$ for different pulse-field amplitudes $\mathcal E_0$. For comparison, we also show the excitation waveform and results obtained in the adiabatic (AA) and diabatic (DA) approximations. 

As  Fig.\ \ref{amplitude_dependence.eps} (a) shows, for field $\mathcal E$ significantly below the adiabatic metallization threshold, $\mathcal E_m=0.75$~V/\protect{\AA}, the AA polarization \cite{Stockman_et_al_PRL_2010_Metallization} and DA population (\ref{diabatic_excitation}) follow the pulse. The computed polarization $\mathcal P$ (green curve)  is close to the adiabatic case except for a small delay and low-amplitude oscillations on the pulse trailing edge with frequency $\approx E_g/\hbar$. This is due to the short duration of the pulse, which leaves at the end a partial coherence between the valence and conduction bands. The population $n_c$ (red curve) is small and dramatically retarded with respect to both the pulse and the DA curve, which is characteristic of the perturbative excitation.

\begin{figure}
\includegraphics[width=.48\textwidth]{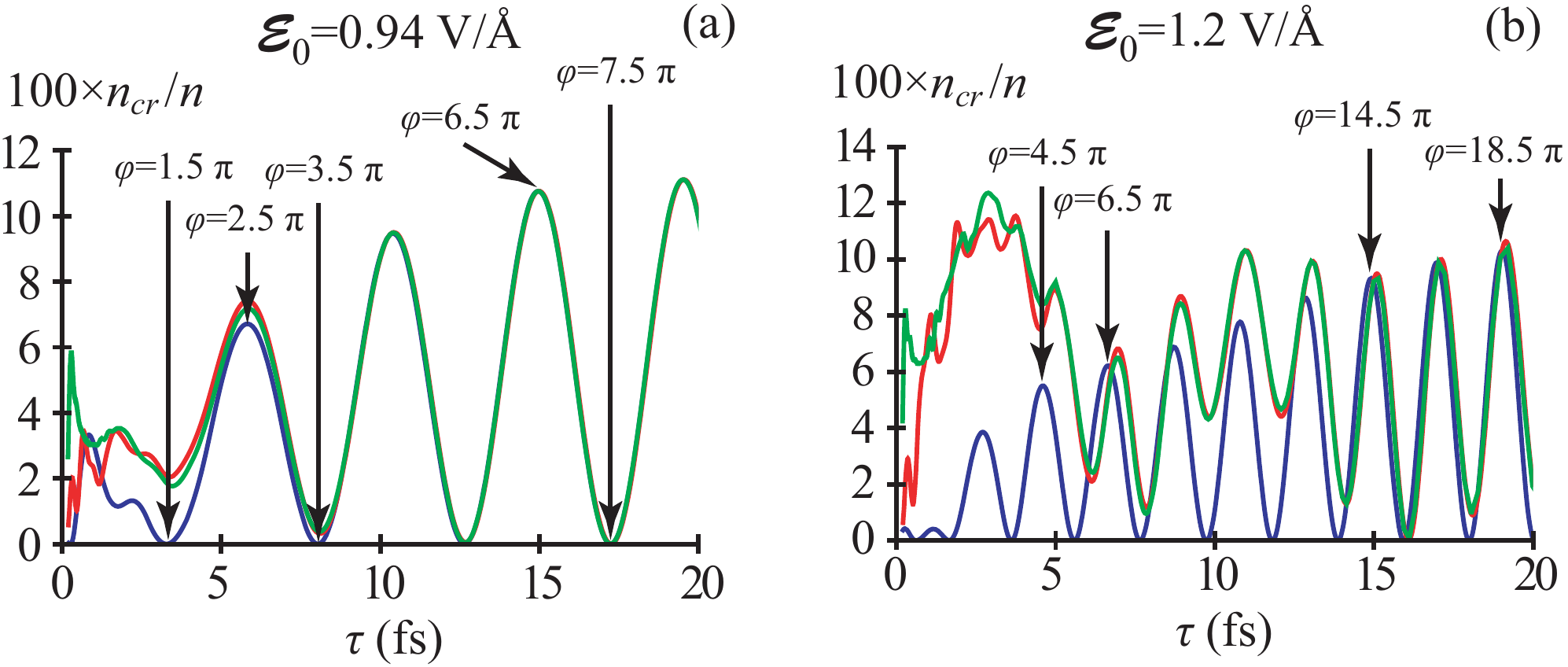}
\caption{Relative residual population of the conduction band $n_{cr}/n$ as a function of the pulse length $\tau$. The green curve is the result of the full numerical computation, the blue curve is the analytical result with two band-edge subbands, and the red with four. Arrows show the adiabatic phase $\varphi$. (a)  $\mathcal E_0=0.94~\mathrm{V/\AA}$ and (b)  $\mathcal E_0=1.2~\mathrm{V/\AA}$.}
\label{dynamics_summary.eps}
\end{figure}

At the threshold of the adiabatic metallization, $E\approx\mathcal E_m$, as illustrated in Fig.\ \ref{amplitude_dependence.eps} (b), the conduction band population $n_c$ dramatically increases. The calculated dependence $n_c(t)$ (red line) agrees well with the DA, except for the residual population after the pulse. The significant deviation of $n_c(t)$  from the DA (gray dash line) starts  at the moment of the second anticrossing on the trailing pulse tail ($t/\tau\approx 0.1$) where the diabatic coupling peaks -- cf.\ Fig.\ \ref{adiabatic_levels.eps} (d). This adiabaticity violation causes the significant residual population $n_{cr}=n_c(t\gg \tau)$ and is dependent on the adiabatic phase $\varphi$ as will be discussed below in conjunction with Fig.\ \ref{dynamics_summary.eps}.

In Fig.\ \ref{amplitude_dependence.eps} (b), the polarization $\mathcal P(t)$ is retarded by an almost quarter pulse length ($\approx \pi/2$ in phase) with respect to the driving pulse, which implies a strong absorption. There also coherent oscillations after the end of the excitation pulse. All this is characteristic of plasmonic metal systems. \cite{Stockman:2002_PRL_control, Stockman_Hewageegana_APA_2007_CEP} The polarization oscillations exhibit beatings between the frequency of the interband and much slower intraband transitions. The latter are caused by the pulse imprinting its frequency by polarizing the hot carriers in the conduction band.  We call this effect  {\em the dynamic metallization}. It is an ultrafast and dissipative  strong-field transition to a plasmonic metal-like behavior.

A similar phase delay between the excitation field and the polarization oscillations has been computed and attributed to the appearance of free electrons in the time-dependent density-functional theory \cite{Otobe_et_al_PRB_2008_Metalization_in_Diamond} of breakdown in bulk dielectrics subjected to high optical fields. Note that such a breakdown for quasi-stationary fields was introduced by Zener. \cite{Zener_Proc_Royal_Soc_1934_Breakdown} Importantly, the present dynamic metallization in thin films is fundamentally different. It is based on the adiabatic contribution to polarization, depends critically on the film thickness,  and occurs at much lower intensities: our field $\mathcal E\lesssim 1$~V/\protect{\AA} corresponds to the radiation intensity $I\lesssim 3\times 10^{13}~\mathrm{W/cm^2}$, in contrast to $I\sim 10^{15}~\mathrm{W/cm^2}$ in Ref.\ \onlinecite{Otobe_et_al_PRB_2008_Metalization_in_Diamond}.

For the 0.85-fs pulse with amplitude $\mathcal E_0=0.94$~V/\protect{\AA}, which is significantly greater than the adiabatic metallization-threshold field $\mathcal E_m=0.75$~V/\protect{\AA} [Fig.\ \ref{amplitude_dependence.eps} (c)], the dynamic metallization phenomena become even more developed. The magnitudes of population $n_c$ and polarization $\mathcal P$ increase. The field time dependence  $n_c(t)$ shows a pronounced saturation between the metallization (anticrossing) points at $t/\tau\approx \pm 0.2$. The residual population forms due to the adiabaticity violation at the anticrossing at $t/\tau\approx 0.2$ and is relatively large because of the large diabatic coupling at this instance -- cf.\ the corresponding (blue) curve in \ Fig.\ \ref{adiabatic_levels.eps} (d). The polarization shows a pronounced plasmonic metal-like behavior: an approximately quarter-oscillation delay with respect to the excitation pulse and the oscillations with a lower frequency, which is close to the pulse mean frequency $\omega_0$.

The excitation dynamics for a longer pulse with $\tau=3.4$ fs is shown in Fig.\   \ref{amplitude_dependence.eps} (d). The main difference from panel (c) is that the polarization $\mathcal P(t)$ peaks almost simultaneously with the excitation pulse maximum, as characteristic of the adiabatic metallization -- cf.\ Ref.\ \onlinecite{Stockman_et_al_PRL_2010_Metallization}. Still there are the residual population $n_{cr}$ and oscillations of $\mathcal P(t)$ after the end of the excitation pulse, which imply non-adiabatic processes occurring at the level anticrossings. 

For the strongly-nonlinear and dispersive problem under consideration, a useful measure of the magnitude of the system's polarizability can reasonably be defined as the effective permittivity 
\begin{equation}
\varepsilon_0=1+4\pi \mathcal P_0/\mathcal E_0~,
\label{epsilon_0}
\end{equation}
 where $\mathcal P_0$ is the maximum value of the polarization in the process. Note that   the maximum of $\mathcal P(t)$ is generally delayed in time with respect to that of $\mathcal E(t)$. For the shorter pulses in Figs.\  \ref{amplitude_dependence.eps} (b)-(c), this delay is approximately $\approx1/4$ of the oscillation length ($\approx \pi/2$ in phase), which implies a strong dissipation (a significant imaginary part of the complex permittivity). 

\begin{figure}
\includegraphics[width=.48\textwidth]{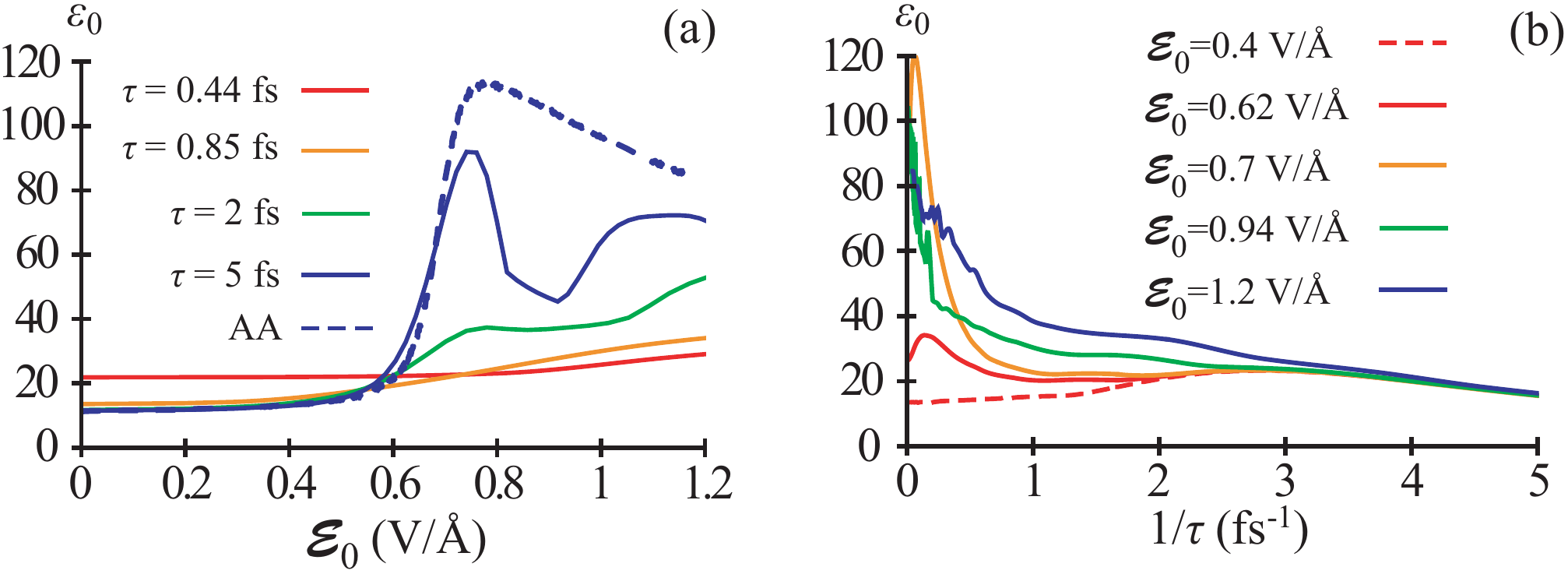}
\caption{
Effective permittivity $\varepsilon_0$ as a function of pulse amplitude $\mathcal E_0$ (a) and inverse pulse duration $1/\tau$ (b). The label AA denotes a result of the adiabatic approximation.
}
\label{epsilon0.eps}
\end{figure}

In Fig.\ \ref{epsilon0.eps} (a)-(b), we plot $\varepsilon_0$ as functions of the excitation-pulse amplitude $\mathcal E_0$ and duration $\tau$. For very short pulses with $\tau\le 0.85$ fs, $\varepsilon_0$ slowly increase with $\mathcal E_0$ due to contribution of perturbative nonlinear absorption. The magnitude $\varepsilon_0\sim 20$  is  rather large because of the wide, high-frequency spectrum of the short pulses. Close to and above the adiabatic metallization threshold, $\mathcal E_0\gtrsim\mathcal E_m=0.75$~V/\protect{\AA},  this effective permittivity dramatically increases for longer pulses with $\tau\gtrsim 2$ fs, suggesting that it is dominated by the adiabatic metallization mechanism. In fact, for $\tau=5$ fs, the dependence $\varepsilon_0(\mathcal E_0)$ resembles that for the adiabatic permittivity \cite{Stockman_et_al_PRL_2010_Metallization} [cf. the blue and dashed blue lines in Fig.\ \ref{epsilon0.eps} (a)]. Note that the appreciable oscillations in the dependence of $\varepsilon_0$ on the pulse duration $\tau$ seen for longer times in Fig.\ \ref{epsilon0.eps} (b) are due to the interference of the excitation amplitudes at the two anticrossings (at the leading and trailing edges of the pulse). These are analogous to the Ramsey fringes, as discussed below for the residual conduction-band population $n_{cr}$ in conjunction with Fig.\ \ref{dynamics_summary.eps}.

In Fig.\ \ref{dynamics_summary.eps}, we display the dependence of the residual (after the pulse end) population $n_{cr}$ on the excitation pulse length $\tau$. A striking feature of this dependence is the presence of high-contrast oscillations. These have a very clear physical origin. In the adiabatic picture, \cite{Stockman_et_al_PRL_2010_Metallization} when the pulse leading-edge field $\mathcal E(t)$ approaches the metallization threshold (causing the anticrossing of the adjacent valence- and conduction-band subbands), the valence electrons are shifted in space to one surface of the nanofilm in the direction of the field where the electrons occupy the quantum-bouncer states. \cite{Stockman_et_al_PRL_2010_Metallization} When the field increases above this metallization threshold, the electron population is coherently transferred to the opposite surface. \cite{Stockman_et_al_PRL_2010_Metallization} This creates polarization oscillating with the transition frequency between the valence- and and conduction-band edges, $\omega_{vc}(t)=\left[E_v(t)-E_c(t)\right]/\hbar$, which then adiabatically evolves with time $t$. The phase accumulated by these oscillations between the time $t_{m1}$ of the anticrossing passage at the leading pulse-edge and that at the trailing edge is $\varphi=\int_{t_{m1}}^{t_{m2}}\omega_{vc}(t)dt$. 

If  $\varphi$ is such that the electrons at the moment $t_{m2}$ are shifted to the initial (in the direction of the maximum pulse field) surface of the nanofilm, then there is a large probability of their return back to the valence band, and the minimum of $n_{cr}$ is observed. Otherwise, the fringe maximum is reached. Thus these oscillations is analogous to the well-known Ramsey fringes. As indicated in Fig.\ \ref{dynamics_summary.eps}, the  adjacent minima and maxima of the $n_{cr}(\tau)$ fringes are indeed separated by the phase change $\Delta\varphi=\pi$. As one can see from Fig.\ \ref{dynamics_summary.eps}, these fringes are described analytically reasonably well with two and very well with four band-edge subbbands  taken into account.

To conclude, in this Letter we have predicted a new effect: ultrafast dynamic metallization of dielectric nanofilms. A single-cycle ultrafast (duration $\sim 1$ fs) optical pulse with the normal electric field of a $\lesssim 1$~V/\protect{\AA} amplitude incident on a dielectric nanofilm (here, a diamond-crystal film with thickness $\sim 2$ nm), induces a plasmonic metal-like dynamics that develops during an ultrashort period on the order of the pulse's duration.  For pulses of $1-2$ fs or longer, this dynamics is characterized by a large, metal-like polarization oscillating with optical frequencies. There is also a  significant residual population of the conduction band, which strongly depends upon and can be coherently controlled by the adiabatic phase $\varphi$ accumulated between the two metalization instances at the leading and trailing edges of the excitation pulse. The polarization oscillations extend beyond the pulse end and also depend on the accumulated adiabatic phase $\varphi$. Thus the dynamic metallization is due to the combination and mutual influence of both the rapid adiabatic (reversible) and diabatic (dissipative) mechanisms.  This dynamic metallization effect can find applications in lightwave electronics, \cite{Goulielmakis_et_al_Lightwave_Electronics_Science_2007} in particular, to create a field-effect transistor controlled by light's electric field with a $\sim 100$ THz bandwidth.

Useful discussions with F. Krausz are gratefully acknowledged. This work was supported by a grant from the Chemical Sciences, Biosciences and Geosciences Division of the BES Office of the Basic Energy Sciences, Office of Science, U.S. Department of Energy,  and by grants from the U.S.--Israel Binational Science Foundation, BaCaTec, and the German Research Foundation (DFG) via the Emmy-Noether program and the Cluster of Excellence: Munich Center for Advanced Photonics.


\end{document}